\documentclass[prl,twocolumn,showpacs,epsfig,longeq]{revtex4}
\usepackage{graphicx}
\usepackage{dcolumn}
\usepackage{amssymb}
\usepackage{bm}
\usepackage{epsfig}
\usepackage{psfrag}

\def\3dots{\:\raisebox{-0.5ex}{$\stackrel{\textstyle.}{:}$}\:}
\def\beq{\begin{equation}}
\def\eeq{\end{equation}}
\def\bea{\begin{eqnarray}}
\def\eea{\end{eqnarray}}

\begin{document}

\title{Probing the doping in metallic and semiconducting carbon nanotubes by Raman and transport measurements}

\author{Anindya Das,$^{1}$ A.K. Sood, $^{1} $\thanks{Corresponding author: asood@physics.iisc.ernet.in} A. Govindaraj, $^{2}$ A. Marco Saitta, $^{3}$ Michele Lazzeri, $^{3}$\\
Francesco Mauri $^{3}$ and C.N.R Rao $^{2}$}

\affiliation{$^{1}$Department of Physics, Indian Institute of Science, Bangalore 560012, INDIA\\
$^{2}$Chemistry $\&$ Physics of Materials Unit, Jawaharlal Nehru Centre for Advanced Scientific Research, Bangalore-560064, India\\
$^{3}$IMPMC, Universit\'es Paris 6 et 7, CNRS, IPGP, 140 rue de Lourmel, 75015 Paris, France}

\date{\today}
\begin{abstract}
In-situ Raman experiments together with transport measurements
have been carried out on carbon nanotubes as a function of gate
voltage. In metallic tubes, a large increase in the Raman frequency
of the $G^-$ band, accompanied by a substantial decrease of its line-width,
is observed with electron or hole doping.
In addition, we see an increase
in Raman frequency of the $G^+$ band in semiconducting tubes.
These results are quantitatively explained using ab-initio calculations
that take into account effects beyond the adiabatic approximation.
Our results imply that Raman spectroscopy can be used as an accurate
measure of the doping of both metallic and semiconducting nanotubes.

\end{abstract}

\maketitle

    Single wall carbon nanotubes (SWNTs) are one dimensional
nanostructures with fascinating electronic, elastic and chemical
properties~\cite{reichbook,saitobook}, with distinct possibility of applications
in nanoelectronics~\cite{nanoelec}, bio, chemical and flow
sensors~\cite{flow1} etc.  {\it In many of these applications, the determination of the doping level
in SWNTs is a crucial issue}.
Raman spectroscopy is one of the most
important characterization techniques.  Low-energy
features ($\sim$ 100/200 cm$^{-1}$), due to the radial
breathing mode (RBM), are used to determine diameter and
chirality of a SWNT~\cite{reichbook}.  Higher-energy modes, near the
graphite Raman $G$ band ($\sim$ 1580 cm$^{-1}$), are often used to
distinguish between semiconducting and metallic tubes.

    In particular, $G$ bands in semiconducting tubes are termed as
$G^{+}$ ($\sim$1590 cm$^{-1}$) and $G^{-}$ ($\sim$1567 cm$^{-1}$) for
LO (axial) and TO (circumferential) modes~\cite{reichbook,dubay02,kohnanomalynt},
respectively. For metallic tubes, $G^{+}$ ($\sim$ 1580 cm$^{-1}$) and
$G^{-}$ ($\sim$ 1540 cm$^{-1}$) peaks are due to TO
(circumferential) and LO (axial) modes~\cite{dubay02,epc,kohnanomalynt},
the opposite of the semiconducting case.
Because of the electron-phonon coupling (EPC) interaction~\cite{epc}, the
$G^-$ linewidth in metallic tubes is usually broader ($\sim$ 60 cm$^{-1}$)
than that in semiconducting tubes ($\sim$ 10 cm$^{-1}$).


\begin{figure}[htp]
\includegraphics[width=0.475\textwidth]{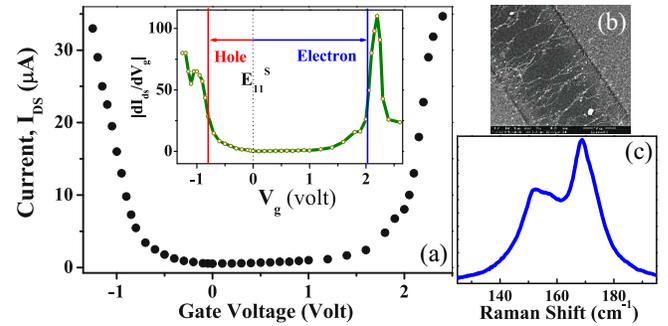}
\caption{(color online). (a) I$_{DS}$ vs Gate voltage (V$_{g}$). The inset shows the derivative of current as a
function of V$_{g}$. The vertical lines (red and blue) correspond to the 1st van-Hove singularities of semiconducting tubes. (b) The SEM picture of our nanotube based FET. The nanotube bundles are connected between two gold electrodes (left and right side). (c) RBM of nanotube bundles at 1.96~eV showing peaks at 151 cm$^{-1}$($d$=1.6 nm) and 170 cm$^{-1}$($d$=1.4 nm).}
\label{Figure1}
\end{figure}

    The phonons associated with
the Raman $G$ band in graphene and with the $G^-$ in metallic SWNTs
are affected by an important Kohn anomaly (KA)~\cite{kohnanomaly,dubay02,kohnanomalynt,epc}.
In graphene, because
of the KA, the $G$ band undergoes a measurable upshift by changing
the Fermi-level with a field-effect device~\cite{lazzeri06,pisana07}.
That is, in graphene, the $G$-band frequency can be used to measure
the actual doping.  Since the origin of the KAs in
graphene and metallic SWNTs are very similar, one could expect
similar effects in metallic SWNTs~\cite{kohnanomalydopednt}. Some results
on the electrochemical doping of nanotubes are already available.
In semiconducting tubes, it
has been seen that the $G^+$ modes shift by a small
amount $\sim$ 1.5 cm$^{-1}$$/$V ~\cite{sem1,semmet1} to higher (lower)
frequencies on hole (electron) doping. In metallic tubes,
the Raman $G^-$ modes harden for both electron and hole doping
by 7 cm$^{-1}$ for a V$_{g}$ of -1 and +1 V~\cite{met1,met2,met3}.
Ref.~\cite{rafailov} qualitatively attributed
the frequency shift of metallic tubes to the KA.
However, the shift of the Fermi level, a key parameter to
understand the results, so far has not been quantified.

\begin{figure*}[htbp]
\includegraphics[width=\textwidth]{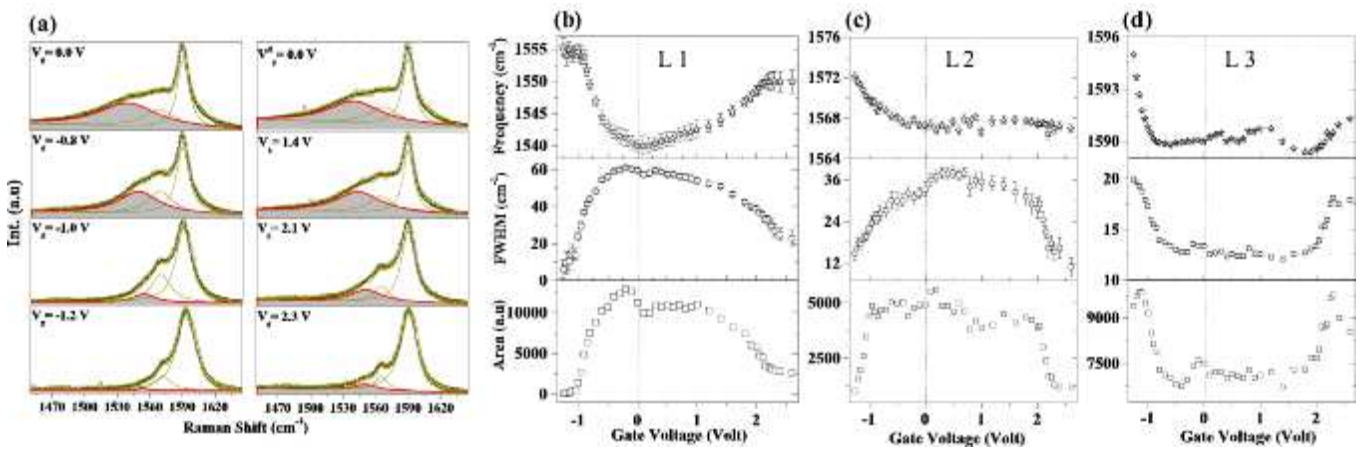}
\caption{(color online). (a) Tangential Raman modes of SWNTs recorded
using excitation energy of 1.96~eV at several V$_g$. The
open circles (dark yellow) shows the raw Raman spectra and the
black lines are the fitted one. The shaded Raman spectra show the
LO phonon component of the metallic nanotubes. V$_g$ dependence of (b) L1, (c) L2 and (d) L3 modes: Frequency (top panel), FWHM (middle panel) and total integrated area (bottom panel).}
\label{Figure2}
\end{figure*}

In this work, we demonstrate that the shift of the $G$ bands can be
used to measure the doping level not only in metallic but also in semiconducting
nanotubes.
This is done by performing a combined Raman and
electron-transport experiments and by comparison with ab-initio
calculations based on density functional theory (DFT).
We use electrochemical gating of the
nanotubes with a solid polymer electrolyte to shift the Fermi level,
which is quantified by electron-transport measurements. This allows a shift of the
Fermi level at much smaller gate voltages as compared to SiO$_{2}$
gating~\cite{electrolyte,electrolyte1,electrolyte2}.

    SWNTs are synthesized by arc-discharge followed by
purification ~\cite{sample}.
AC dielectrophoresis~\cite{dielectrophoresis} is used to align and
connect the nanotubes between two gold electrodes at 15 $\mu$m
separation. The average diameter of the bundles connected
to the electrodes is 10-50 nm from the SEM image, Fig. 1(b).
We have used as gating material a solid-polymer
electrolyte~\cite{electrolyte1,electrolyte2} prepared by dissolving
LiClO$_{4}$ and PEO (1:0.12) in methanol to form a precursor. The
gate is applied by placing a platinum electrode into the polymer
layer. The advantage of the solid electrolyte over the
electrolytes in solution phase is that it does not degrade the
sample and electrodes and the gate-current is extremely small
($\sim$nA) even at high gate voltages ($\sim$1V). A DILOR triple
grating spectrometer equipped with charge-couple device is used to
record the Raman spectra using a He-Ne (1.96~eV) laser.
Fig. 1(c) shows the RBM of our tubes. It has two bands with average diameter $d$=1.4 and 1.6 nm. The laser at 1.96~eV is in resonance with E$^M$$_{11}$ of tube diameter 1.4 nm and lies in between the E$^M$$_{11}$ and E$^S$$_{33}$ of 1.6 nm diameter tubes~\cite{kataura}. Therefore, Raman spectra has contributions from metallic tubes of $d$=1.4 nm as well as from both semiconducting and metallic tubes of $d$=1.6 nm.

    Fig. 1(a) shows the ambipolar behavior of our nanotube FET
device at a drain-source voltage V$_{DS}$ of 50~mV. At gate voltage
V$_g$=0, I$_{DS}$ $\sim$ 500 nA. The behavior of the current through
the SWNT is reversible with V$_g$ and the leakage current is very
small ($\sim$ 10 nA at V$_g \sim$ 2V).  Since the current increases on
either side of the starting zero V$_g$, we infer that the initial
Fermi energy ($\epsilon_{\rm F}$=0) is at the middle of the gap of
semiconducting tubes (charge neutrality point-CNP)~\cite{electrolyte2}.
The sharp increase in current, as shown in Fig. 1(a), for large positive (negative) V$_g$ is due
to addition of electrons (holes) to the first van Hove singularity on conduction (valence) band side of the semiconducting tubes present in the sample along with the metallic tubes. The
onset of this increase in current is observed for V$_g$=2.02 V and
-0.8 V, inset of Fig. 1(a). For the semiconducting tubes of
average diameter $\sim$ 1.5 nm, E$^S$$_{11}$ $\sim$ 0.5~eV and for
metallic tubes E$^M$$_{11}$ $\sim$ 1.8~eV~\cite{kataura}. As the
metallic tubes have larger energy separation between the van Hove
singularities, the sharp increase in conductivity in Fig. 1(a) is
mostly governed by the semiconducting tubes. Therefore, the semiconducting tubes of the sample act as an
internal read out to estimate the V$_g$-induced $\epsilon_{\rm F}$
shift. We define the proportionality factor
$\alpha_e$ ($\alpha_h$), to estimate $\epsilon_{\rm F}$ at
different positive (negative) gate voltages ($\epsilon_{\rm F}$ =
$\alpha$V$_{g}$). For electron doping, a positive voltage of
$\sim$ 2.02~V shifts $\epsilon_{\rm F}$ into the first conduction
band, which is 0.25~eV from the CNP. Therefore,
$\alpha_e$ = 0.25/2.02 = 0.12~$e$ and similarly $\alpha_h$ =
0.25/0.8 = 0.31~$e$. The difference between
$\alpha_e$ and $\alpha_h$ is probably due to double
layer formation of different ions Li$^{+}$ and ClO$_{4}$$^{-}$ in PEO
matrix at the interfaces.

    Fig. 2(a) shows the Raman spectra recorded at different V$_g$. A
constant gate voltage is applied for 15 minutes to stabilize I$_{DS}$
and then the Raman spectra were recorded for next 15 minutes at the
same V$_g$. For a V$_g$ in the -1.25/+2.5 V range, the Raman spectra
as well as the source-drain current fully recover to the starting
condition (V$_g$=0) within about one hour after removing the gate voltage.
There are three prominent Raman modes and the spectra
are fitted with three Lorentzians L1 ($\sim$1540 cm$^{-1}$), L2
($\sim$1567 cm$^{-1}$) and L3 ($\sim$1590 cm$^{-1}$). The Raman frequency, linewidth and total area thus obtained for L1, L2 and L3
modes are plotted in Figs. 2(b), 2(c) and 2(d), as a function of V$_g$. Following Fig.19 of Ref.~\cite{kohnanomalynt},
the L1 line is attributed to the LO mode of the metallic tubes with diameter 1.4
nm~\cite{kohnanomalynt}. The L1 line can be equally well fitted by a
Lorentzian or a Fano-resonance line-shape. We use a Lorentzian line-shape to
minimize the number of fitting parameters. In our
spectra, the L2 has a large line-width of $\sim$ 40 cm$^{-1}$ and the doping dependence is similar to that of L1 mode. Therefore the L2 line is attributed to a combination of the LO mode of metallic tubes and of the TO mode of semiconducting tubes with diameter 1.6 nm (Fig.19 of Ref.~\cite{kohnanomalynt}). The L3 line is attributed to a combination of the TO of metallic and the LO of semiconducting tubes. The EPC determines an important broadening of the LO mode of metallic
tubes~\cite{epc}. This explains the large width of the L1 ($\sim$
60~cm$^{-1}$) and L2 ($\sim$ 40~cm$^{-1}$) lines. The sharpness of the
L3 line ($\sim$ 10~cm$^{-1}$) is explained considering that the
semiconducting LO and the metallic TO modes are not broadened by
EPC and that their frequencies are almost independent from diameter
(Fig.19 of Ref.~\cite{kohnanomalynt}).

    As shown in Fig. 2(b), the largest effect is the frequency
increase of the L1 line by 15/10 cm$^{-1}$ on hole/electron doping. The
substantial decrease of its linewidth is also remarkable.  Both
effects are due to the dependence of the KA on V$_g$ in
metallic tubes.  The L2 line displays a similar behavior due to
the metallic tubes associated with this line.
The intensity of the L1 line (entirely due to metallic tubes)
drops to zero at high doping (positive or negative). This suggests that,
at high doping, the Raman signal from metallic tubes is neglegible.
Thus, the behavior of the L3 line at high doping is mostly
due to semiconducting tubes.
Interestingly, the frequency upshift of the L3 line is
perfectly correlated to the I$_{DS}$ current (Fig.1), indicating that
the L3 shift is due to the change of the
electron (hole) populations in the conduction (valence) bands of
semiconducting tubes.

We now compare measurements with DFT calculations, done using i)
the adiabatic Born-Oppenheimer approximation (BOA, Eq.6 of
Ref.~\cite{lazzeri06}) and ii) time-dependent perturbation theory
(TDPT, Eq.7 of Ref.~\cite{lazzeri06}) to include dynamical effects
beyond BOA.  Within BOA, the phonon frequency is obtained from the
forces resulting from a {\it static} displacement of the atoms.
However, a phonon is not a static perturbation to the system but a
{\it dynamic} one, oscillating in time, that should be treated within
TDPT.  Dynamic effects beyond BOA are usually assumed to be
neglegible, but they are crucial to reproduce the Raman spectra of
nanotubes and doped
graphene~\cite{kohnanomalynt,kohnanomalydopednt,lazzeri06,pisana07}.
Our DFT calculations are based on a linear response
approach~\cite{dfpt} as described in Ref.~\cite{kohnanomalydopednt}.
Calculations are done on a metallic (10,10) and on a semiconducting
(20,0) tube, having diameters 1.4 and 1.6~nm.  We compute the
variation of the LO and TO phonon frequency ($\Delta\omega_{\rm LO}$
and $\Delta\omega_{\rm TO}$) as a function of the doping.  We also
compute the intrinsic linewidth ($\gamma_{\rm LO}$ and
$\gamma_{\rm TO}$) due to EPC (Eq.29 of
Ref.~\cite{kohnanomalydopednt}), owing to the decay of the phonon into
electron-hole pairs, which is nonzero only for metallic tubes.
The electronic occupations are determined according to a
Fermi Dirac distribution at 315~K.

\begin{figure}[tbp]
\includegraphics[width=0.4\textwidth]{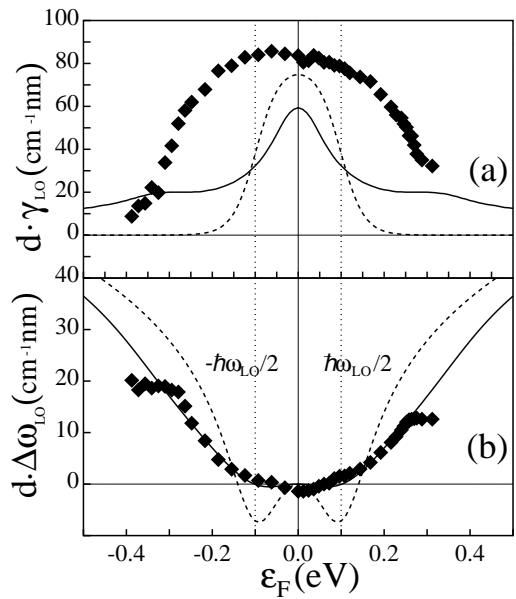}
\caption{
$G^{-}$ peak in metallic tubes: (a) linewidth
($\gamma_{\rm LO}$, FWHM) and (b) frequency shift ($\Delta\omega_{\rm
LO}$) as a function of the doping. The Fermi energy
($\epsilon_{\rm F}$) is zero at the charge neutrality  point.
$\gamma_{\rm LO}$ and $\Delta\omega_{\rm LO}$ are multiplied by
the the tube diameter ($d$) to obtain an universal behavior.
Diamonds are measurements (L1 line). Lines are calculations done considering
(full) or not considering (dashed) an inhomogeneous fluctuation of
$\epsilon_{\rm F}$. $\omega_{\rm LO}$ is the frequency of the LO
phonon.  \label{Figure3}}
\end{figure}

For metallic tubes $\Delta\omega_{\rm TO}$ and $\gamma_{\rm TO}$
are negligible, since the EPC between occupied and empty
electronic states is zero (see Eq.9 of
Ref.~\cite{kohnanomalydopednt}).  For diameter $d$ larger than
1~nm, $d\times\Delta\omega_{\rm LO}$, and $d\times\gamma_{\rm LO}$
are universal, i.e.  independent of both $d$ and
chirality~\cite{epc,kohnanomalydopednt}.  Fig.\ref{Figure3}
reports these universal behaviors, as functions of $\epsilon_{\rm
F}$, computed within TDPT. BOA results are not reported, since
they are very similar for $\hbar\omega_{LO}/2<|\epsilon_{\rm
F}|<1.0$~eV~\cite{kohnanomalydopednt}.  The hardening of
$\Delta\omega_{\rm LO}$ for $|\epsilon_{\rm
F}|>\hbar\omega_{LO}/2$ is understood by considering the
modification of the electronic band structure in the presence of
the LO phonon. An atomic distortion following the LO phonon
pattern opens a gap at the CNP~\cite{dubay02}. The filling of the
electronic states with doping increases the energy required to
open this gap, thus inducing a hardening of the LO
frequency~\cite{footnotehardening}.

We assume that $\epsilon_{\rm F}$ has a Gaussian distribution
(with a standard deviation $\sigma$) to take into account the spatial
fluctuation of $\epsilon_{\rm F}$ in the actual device.  In
Fig.~\ref{Figure3} we report the average frequency and the FWHM of the
LO peak profile obtained by an ensemble of tubes with such a
distribution.  To include the experimental data in Fig.~\ref{Figure3}
we multiply the measured V$_g$ by $\alpha_e$ and $\alpha_h$.  The
calculations best describe the experiment with $\sigma=0.055$~eV.  The
agreement between theory and measurements is excellent for
$\Delta\omega_{\rm LO}$.  The theoretical LO linewidth is smaller than
the experimental one, probably because of other effects, such as the
finite diameter distribution.

\begin{figure}[tbp]
\includegraphics[width=0.4\textwidth]{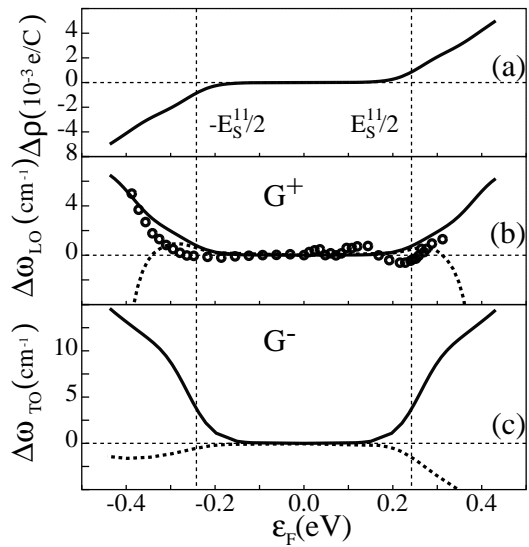}
\caption{$G^{+}$ and $G^{-}$ peaks in 1.6~nm semiconducting tubes:
(a) Theoretical charge doping per C atom. (b,c) Frequency shift
($\Delta\omega_{\rm LO}$, $\Delta\omega_{\rm TO}$) as a function of the
Fermi energy ($\epsilon_{\rm F}$). Open dots are measurements (L3 line).
Dotted lines are adiabatic calculations (BOA).
Full lines are our most  accurate theoretical results
that take into account dynamical effects (TDPT).
Vertical lines are the theoretical position of the first van Hove
singularities. \label{Figure4}}
\end{figure}

In semiconducting tubes, $\Delta\omega_{\rm LO}$ and
$\Delta\omega_{\rm TO}$ depend on the positions of the band edges,
i.e. on $d$ and, to a lesser extent, on the chirality.
Fig.~\ref{Figure4} shows $\Delta\omega_{\rm LO}$ and
$\Delta\omega_{\rm TO}$ computed within BOA and TDPT.  Within the
more accurate TDPT, $\Delta\omega_{\rm LO}$ and $\Delta\omega_{\rm
TO}$ upshift when $\epsilon_{\rm F}$ reaches the vHs.  To
understand this behavior, we have to consider how the electrons
screen a phonon vibration.  The electronic screening can be
expressed in terms of {\it interband} and {\it intraband} EPC
transitions between electronic states~\cite{lazzeri06}.  Within
TDPT only the {\it interband} transitions between occupied valence
and empty conduction electronic states contribute (Eq.(7) of
~\cite{lazzeri06}).  By doping the tube, the valence (conduction)
band is emptied (filled). Both processes correspond to a reduction
of the number of allowed transitions.  The consequent reduction of
the screening upshifts $\omega$.  The shift is more pronounced for
the TO phonon that has a larger EPC for transition beween valence
and conduction band edges. Within BOA, the presence of {\it
intraband} transitions (second line of Eq.~(6) of
~\cite{lazzeri06}) contrasts the hardening due to {\it interband}
transition~\cite{lazzeri06,kohnanomalydopednt}.

The $G^-$ of both metallic and semiconducting tubes are affected in a
different way by V$_g$.  Thus, the behavior of $G^-$ of semiconducting
tubes in the L2 line cannot be separated from that of metallic tubes.
We can instead directely compare the theoretical $\Delta\omega_{\rm
LO}$ with the L3 line, since, at small doping, the position the $G^+$
of both metallic and semiconducting tubes does not depend on
V$_g$ and, at large doping, the metallic-tube intensity
vanishes.  TDPT calculations for the LO mode compare very well with the
mesured behavior of the L3 line, that we report in Fig.~\ref{Figure4}
using the transport-derived ratios $\alpha_e$, $\alpha_h$.  At the
minumum V$_g$, L3 upshifts by $\sim$ 5~cm$^{-1}$. According to
TDPT, this corresponds to a doping of $4.0\times10^{-3}$
holes per C atom and to -0.15~eV shift of $\epsilon_{\rm F}$ below the
first vHs.

Concluding, by combining transport, in-situ Raman experiments
and ab-initio calculations we quantified the effect of the doping on
the phonons in carbon nanotubes.  By electron and hole doping, the
Raman $G^-$ peak of metallic tubes and both the $G^-$ and $G^+$ of
semiconducting tubes harden. Moreover, the $G^-$ linewidth of metallic
tubes narrows.  Important consequences of the present results are
that: i) The metallicity of the system is not always associated with
the presence of a broad $G^-$ peak, as often assumed. Indeed, in a
doped metallic tube, the $G^-$-peak width can be much smaller than in
the undoped case if the Fermi level is far from the $\pi$
band-crossing.  ii) Given the strong dependence of the Raman $G$ bands
on the Fermi energy, Raman spectroscopy can be used as an accurate
measure of the doping of both metallic and semiconducting nanotubes,
with important technological implications for nano electronics.

    AKS thanks DST, India for support. Calculations were performed at IDRIS
(France), project CP9-61387/71387, using the quantum-espresso package
(www.pwscf.org).


\begin{thebibliography}{10}
\bibitem{footnotehardening}
{The behavior of $\Delta\omega_{\rm LO}$ in metallic tubes can be
equally understood within perturbation theory, in terms of
interband electronic transitions, as we explained in
ref.~\cite{kohnanomalydopednt}. Here we adopt this approach to
interpret the behavior of the semiconducting tubes.}
\end{thebibliography}
\end{document}